\documentclass{cernyrep2}

\begin{document}
\let\TT\textsuperscript

\title{Track Based Alignment in CMS\footnote{Contributed to the proceedings of the 1st LHC Detector Alignment Workshop, CERN, September 2006}}
\author{F.-P.~Schilling\TT{a}\\
        for the CMS collaboration\\[14pt]
\TT{a} CERN, Geneva, Switzerland\\
\texttt{Frank-Peter.Schilling@cern.ch}}



\maketitle

\begin{multicols}{2}


\begin{abstract}

The strategy for track based alignment of the CMS tracking and muon
detectors is presented. After an overview over the used data samples,
the general alignment strategy is presented, with a focus of the
procedures envisaged at the start of data taking in 2007/8. The 
three currently used alignment algorithms are discussed and first
results on their application to the CMS tracker are presented, as well
as studies on the alignment of the Muon detector with tracks.

\end{abstract}


\section{Introduction}

The alignment of the CMS tracking and muon detectors represents a
particular challenge, because the number of alignment parameters to be
determined with high accuracy is very large.  In particular, aligning
the $\sim 15,000$ silicon modules of the pixel and strip tracker with
a precision which is comparable to or better than their intrinsic
resolution of $10\ldots50 \rm\ \mu m$, requires solving a problem with
$\mathcal{O}(100,000)$ unknowns.  In addition to survey measurements
at construction time and optical (laser) alignment~\cite{villa,arce}, 
track based
alignment will be a necessary ingredient to approach this huge
challenge (see also~\cite{webertalk}).

At the time of this workshop, a lot of activities related to track
based alignment were ongoing already in CMS, and are summarized
in~\cite{ptdr1}.  Three different alignment algorithms have been
implemented in the CMS software within a common framework. Alignment
studies applying these algorithms to various Monte Carlo data sets
have been performed. The advantage of having several algorithms at
disposal is that the obtained results can be cross-checked using
different algorithms with different systematics. See
section~\ref{sec:datasets} on the data sets considered for alignment.
Other activities related to track-based alignment are the development
of a dedicated alignment stream produced during the prompt
reconstruction at the Tier-0, which uses a special reduced data format
(``AlCaReco''). It will enable to run track based alignment with a
short turn-around time after the data have been taken. Moreover, a
framework is being developed in order to combine the results of
track-based and laser alignment, and with survey data. The benefits of
using mass and vertex constraints, as well as of overlapping tracker
modules are also studied.  Further work is ongoing in order to
establish observables sensitive to misalignment other than $\chi^2$,
in order to fix global transformations which leave the $\chi^2$
unchanged.


\section{Data Samples}
\label{sec:datasets}

The following data samples are currently considered for alignment:

\begin{itemize}

\item {\bf High $\mathbf{p_T}$ muons from $\mathbf{Z,W}$:} \\
These constitute the primary source of high quality tracks for
alignment, because of their high transverse momentum and small amount
of multiple scattering in the tracker material.  A compilation of the
expected event rates after HLT is shown in Tab.~\ref{tab:zwrates}.

\item {\bf Cosmic muons:} \\
Cosmics are a valuable source of tracks since they are available
already before the first beams are in the LHC machine. They are
particular useful for the alignment of the barrel tracker and muon
detectors.  Estimates show that the rate for cosmic muons accepted by
the L1 trigger and standalone muon reconstruction is around $\sim400
\rm\ Hz$~\cite{cosmicnote}.

\item {\bf Beam halo muons:} \\
As soon as the LHC will be commissioned with single beams,
near-horizontal beam halo muons constitute a valuable source of tracks
for the alignment of the tracker and muon endcaps.  A rate of $\sim5
\rm\ kHz$ is expected per side accepted by the L1 muon trigger and
standalone muon reconstruction~\cite{cosmicnote}.  A problem arises
for the tracker due to the fact that the muon endcap trigger has no
acceptance for radii covered by the tracker endcap.  It is foreseen to
install dedicated scintillators in order to trigger beam halo muons
within the tracker acceptance.

\item {\bf Muons from $\mathbf{J/\Psi}$ and b hadron decays:} \\
Even though having a comparatively small transverse momentum spectrum,
these will be very useful in particular at the start-up of the LHC,
when luminosities will be modest and muons from $Z,W$ decays are not
yet available.  In addition, for muons from $J/\Psi\to\mu^+\mu^-$ an
invariant mass constraint can be used.

\item {\bf Isolated tracks from QCD events:} \\
At low luminosities this will be the only source of collision tracks
which might be useful for alignment.  Multiple scattering in the
tracker material is clearly an issue, but studies show that these
events could be useful at least for the alignment of the 
pixel detector.

\end{itemize}


\section{Alignment Strategy}

The earliest information on the tracker and muon alignment will come
from survey measurements carried out during construction, as well as
from the laser alignment systems.  The alignment can then be improved
upon with cosmics as well as with beam halo muons. The current LHC
startup schedule foresees a 2-3 week {\em calibration run} at
$\sqrt{s}=900 \rm\ GeV$ and luminosities not exceeding $10^{29} \rm\
cm^{-2}s^{-1}$ at the end of 2007. Since no high $p_T$ muons from
$W,Z$ decays will be available under these conditions, minimum bias
QCD events will constitute the only source of tracks for alignment. In
addition, the pixel detector will not yet be installed. If possible, a
first track-based alignment of composite strip tracker structures
(layers etc.) could be performed.

During the winter shutdown 2007/8 the pixel detector will be installed and
the subsequent {\em pilot physics run} will aim at luminosities up to
$10^{32} \rm\ cm^{-2}s^{-1}$. This will allow to accumulate large
statistics of high quality muon tracks from $W,Z$ decays in a short
timescale (Table~\ref{tab:zwrates}). A two-step procedure is currently
foreseen to align the tracker: First, a standalone alignment of the
pixel detector will be carried out. Second, the strip tracker will be
aligned, using the pixel detector as a reference system.


\section{Alignment Algorithms}

Track-based alignment was shown to be the optimal method for the
alignment of large tracking detectors in previous
experiments. However, it represents a major challenge at CMS because
the number of degrees of freedom involved is very large: Considering
3+3 translational and rotational degrees of freedom for each of the
$\sim 15000$ modules leads to $\mathcal{O}(100,000)$ alignment
parameters, which have to be determined with a precision of $\sim 10
\rm\ \mu m$. Moreover, the full covariance matrix is of size
$\mathcal{O}(100,000 {\rm x} 100,000)$.

In CMS, three different track-based alignment algorithms are
implemented in the reconstruction software, using a common software
framework.  Some of them have been successfully used at other
experiments, others are newly developed. In the following, the main
features and initial results of using these algorithms in CMS are
summarized.


\subsection{General Software Framework}

Within the CMS software it is not necessary to apply (mis)alignment
corrections to the geometry already at the simulation step. Instead,
(mis)alignment can be applied ``on the fly'' during
reconstruction. Dedicated software tools have been implemented to move
and rotate parts of the tracking or muon detectors in a hierarchical
way~\cite{misaliscen}.  In addition, a so-called {\em alignment
position error} can be added to the intrinsic uncertainty of
reconstructed hits in order to take the effects of misalignment into
account in the track reconstruction. Two dedicated {\em misalignment
scenarios}~\cite{misalinote} have been implemented which emulate the
expected misalignment for different phases of data taking: the {\em
First Data Taking} scenario and the {\em Long Term} scenario (for
details see~\cite{steinbrueck}).
A fast track refit has been implemented in the reconstruction
software, such that redoing the full pattern recognition is
avoided~\footnote{The assumption that misalignment does not change the
assignment of hits to tracks was verified for the case of not too
large misalignments.}.

Alignment studies are performed using the reduced {\em AlCaReco}
format, in which only the tracks used for alignment are kept in the
event (e.g. the two muon tracks in case of $Z^0\to\mu^+\mu^-$
events). This significantly improves both the disk space needed as
well as the alignment algorithm performance.

The alignment algorithms have been implemented in the standard CMS
reconstruction software using a common layer of software, which
provides all features which are common to all algorithms, for instance
the management of alignment parameters and covariance matrices, the
calculation of derivatives with respect to track or alignment
parameters, input/output and an interface to the CMS offline
conditions database.


\subsection{HIP Algorithm}

An iterative alignment algorithm using the Hits and Impact Points
(HIP) method was developed in~\cite{hiplajolla}. It is able to
determine the alignment of individual sensors by minimizing a local
$\chi^2$ function depending on the alignment parameters, constructed
from the track-hit residuals on the sensor.  Correlations between
different sensors are not explicitly included, but taken care of
implicitly by iterating the method, which involves consecutive cycles
of calculating the alignment parameters and refitting the tracks. The
algorithm is computationally light because no inversion of large
matrices is involved.  An alternative implementation of the algorithm
is designed to align composite detector structures for a common
translation and rotation~\cite{hipnote}, for example pixel ladders or
layers. The composite alignment involves only a small number of
parameters, and therefore a rather small number of tracks is
sufficient to carry out alignment already in the beginning of data
taking.

\begin{figure}
\centering
\begin{minipage}{0.65\linewidth}
\includegraphics[angle=270,width=0.9\linewidth]{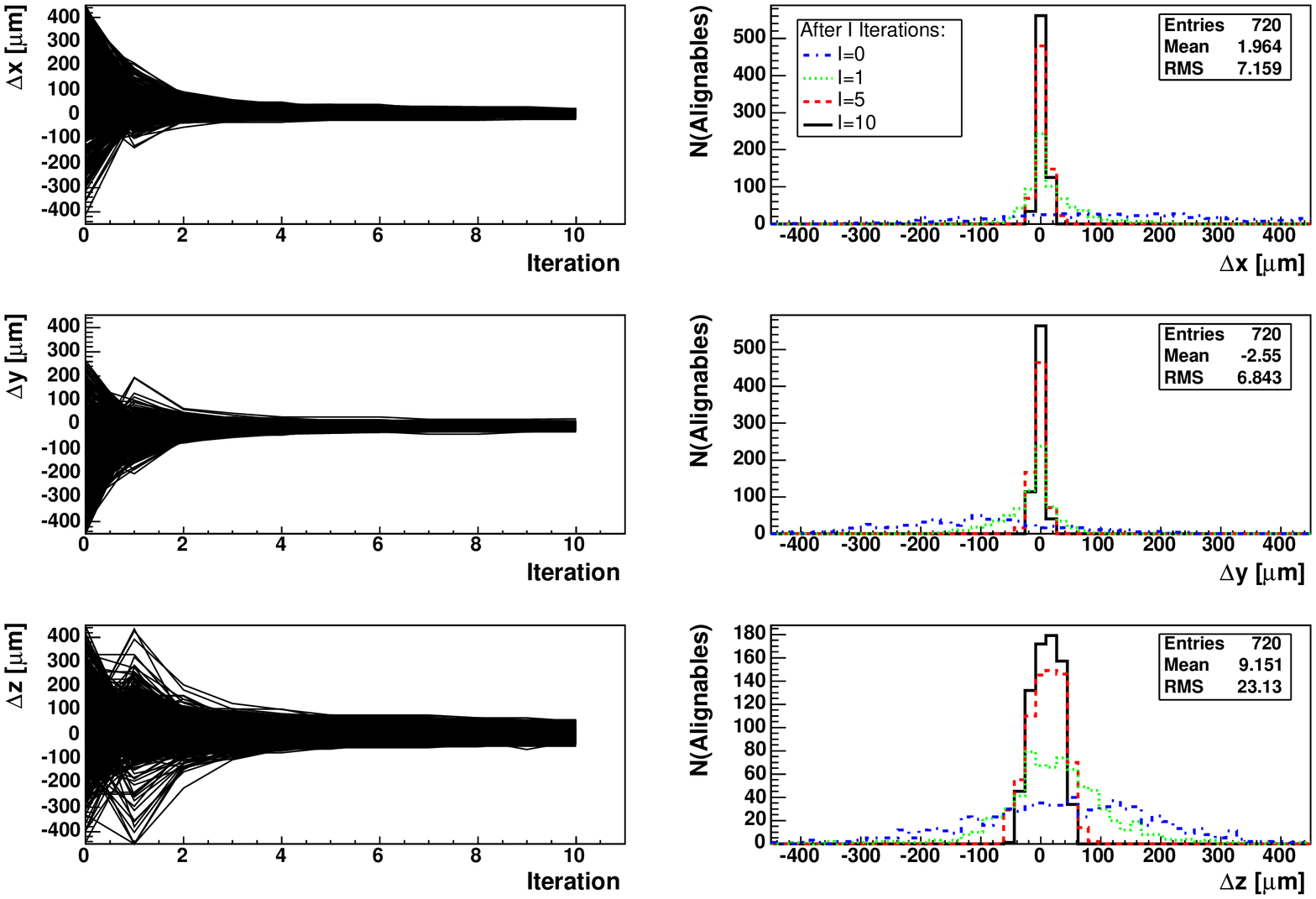}
\end{minipage}
\begin{minipage}{0.32\linewidth}
\includegraphics[width=0.9\linewidth]{Figure_006-028-a.eps} \\
\includegraphics[width=0.9\linewidth]{Figure_006-028-c.eps}
\end{minipage}
\caption{
Left: Alignment of the Pixel barrel modules with the HIP algorithm.
The residuals in global coordinates are shown as a function
of iteration, and projected for 0,1,5 and 10 iterations. Right:
Right: Kalman Filter alignment. Residuals in local x for TIB layers 1 (top)
and 2 (bottom) as a function of the number of processed tracks. }
\label{fig:hipkalman}
\end{figure}

The HIP algorithm has been used in~\cite{hipnote} for the alignment of
the pixel barrel modules using the First Data Taking misalignment
scenario.  The pixel endcaps and the strip tracker are not misaligned.
The procedure has been iterated 10 times using 200 000 simulated
$Z^0\rightarrow\mu^+\mu^-$ events. Figure~\ref{fig:hipkalman} (left)
shows the differences between the true and estimated alignment
parameters. The convergence is good, with RMS values of $7(23)\rm\ \mu
m$ for the $x,y(z)$ coordinates, respectively. The algorithm was also
applied to a test beam setup~\cite{hipcrack}.


\subsection{Kalman Filter Algorithm}

A method for global alignment using charged tracks can be derived from
the Kalman filter. The method is iterative, so that the alignment
parameters are updated after each track. It can be formulated in such
a way that no large matrices have to be inverted~\cite{kalmannote}. In
order to achieve a global alignment the update is not restricted to
the detector elements that are crossed by the track, but can be
extended to those elements that have significant correlations with the
ones in the current track. This requires some bookkeeping, but keeps
the computational load to an acceptable level.  It is possible to use
prior information about the alignment obtained from mechanical survey
measurements as well as from laser alignment. The algorithm can also
be extended to deal with kinematically constrained track pairs
(originating from particle decays).

The algorithm has been implemented in the CMS software and studied in
two small subsets of the silicon tracker: A telescope-like section of
the inner and outer barrel, and a wheel-like subset of the inner
barrel, consisting of 156 modules in 4 layers. The tracks used were
simulated single muons with $p_T=100 \rm\ GeV$. Random misalignment
with a standard deviation of $\sigma=100 \rm\ \mu m$ was applied to
the local $x$ and $y$ positions of the modules.  Results from the
alignment of the wheel-like setup are shown in
Figure~\ref{fig:hipkalman} (right).  It shows the evolution of the
differences between true and estimated $x$-shifts for layers 1 and
2. A total of 100 000 tracks were processed. As can be seen, the speed
of convergence depends on the layer. For more details,
see~\cite{fruehwirth}.


\subsection{Millepede-II Algorithm}

Millepede~\cite{blobel} is a well established and robust program
package for alignment which has been used successfully at other
experiments, for example at H1, CDF, LHCb and others.  Being a
non-iterative method, it has been shown that it can improve the
alignment precision considerably with respect to other algorithms.

Millepede is a linear least-squares algorithm which is fast, accurate
and can take into account correlations among parameters. In the
least-squares fit local track parameters and global alignment parameters
are fitted simultaneously.  The solution for the alignment parameters
is obtained from a matrix equation for the global parameters only.
For $N$ alignment parameters this requires the inversion of a $N {\rm
x} N$ matrix. However, this method can only be used up to $N\sim
10000$ due to CPU and memory constraints.  The alignment of the CMS tracker
exceeds this limit by one order of magnitude. Therefore, a new version
Millepede-II~\cite{millenote} was developed, which offers different
solution methods, and is applicable for $N$ much larger than $10000$.
In Millepede-II, in addition to the matrix inversion
and a diagonalization method, a new method for the solution
of very large matrix equations is implemented. This minimum
residual method applicable for sparse matrices determines
a good solution by iteration in acceptable time even for large $N$.
For more details, see~\cite{blobeltalk}.

\begin{figure}
\centering
\includegraphics[width=0.45\linewidth]{Figure_006-029-b.eps}
\hfill
\includegraphics[width=0.45\linewidth]{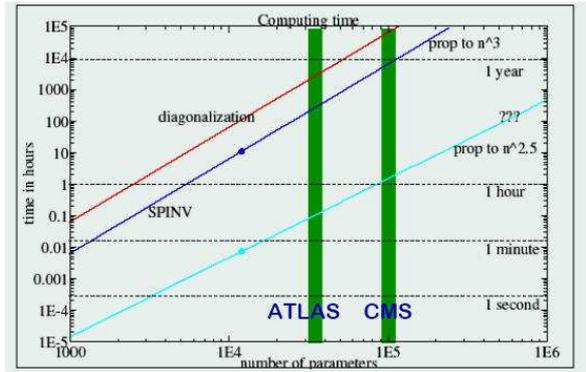}
\caption{
Millepede-II: Left: Residuals in $r\phi$ in the strip tracker barrel before (red) and
after (black) alignment using Millepede-II.
Right: CPU time as a function of alignment parameters for matrix
inversion (blue) and Millepede-II.
}
\label{fig:millepede}
\end{figure}

Millepede-II has been interfaced to the CMS software and the
alignment of parts of the CMS tracker has been carried out using
different scenarios~\cite{millenote}.  As an example,
Figure~\ref{fig:millepede} (left) shows hit residuals in $r\phi$ for
the new iterative method. Each individual sensor of the tracker was
misaligned.  The alignment procedure was carried out in the barrel
region ($|\eta|<0.9$) of the strip tracker using 1.8 million
$Z^0\rightarrow\mu^+\mu^-$ events.  The pixel layers and the outermost
barrel layer were kept fixed, resulting in $\sim 8400$ alignment
parameters. The convergence is very good, and the results obtained are
identical to those using the matrix inversion method, but the new
method being faster by about three orders of magnitude.

Figure~\ref{fig:millepede} (right) shows the needed CPU time as a
function of the number of alignment parameters for the diagonalization
and matrix inversion methods as well as for the new method used in
Millepede-II.  It can be seen that Millepede-II is expected to be
capable to solve the full CMS tracker alignment problem within
reasonable CPU time.

Millepede-II has also been used~\cite{stoye} to investigate the global
correlations between alignment parameters in the case of the CMS
tracker. It turns out that in certain cases these correlations can be
very high (above $99\%$). Studys show that it is very important to combine
samples of tracks with different topologies, such as collision tracks
and cosmics, in order to reduce these global correlations.


\section{Muon Alignment}

The CMS Muon system consists of 790 individual chambers with an
intrinsic resolution in the range $75\ldots100 \rm\ \mu m$.  Excellent
alignment of the muon system is particularly important to ensure
efficient muon triggering and good track momentum resolution at large
momenta, where the resolution is dominated by the muon detector.

For optimal performance of the muon spectrometer over the entire
momentum range up to 1 TeV, the different muon chambers must be
aligned with respect to each other and to the central tracking system
to within $100\ldots500 \rm\ \mu m$. To control misalignment during
commissioning and to monitor further displacements during operation,
which can be of the order of one mm, CMS will combine measurements
from an optical-mechanical system with the results of track based
alignment~\cite{muonalinote}. Two approaches are pursued: alignment
using tracks which are extrapolated from the tracker, and a standalone
muon alignment (Fig.~\ref{fig:muonali}).


\section{Conclusions}

The alignment of the CMS tracker and muon detectors constitutes a
significant challenge due to the large number of parameters ($\sim
100,000$ in the tracker) as well as the high intrinsic resolution of
the detectors.

Even though LHC operation is still more than one year away, a lot of
activities are ongoing already now related to track based alignment in
CMS. The initial results obtained with the three considered alignment
algorithms are very promising, although a realistic alignment of the
full tracker at the sensor level is yet to be demonstrated. In
addition, real data from tracker test setups as well as from the {\em
Magnet Test and Cosmic Challenge} are being studied for
alignment. Work is ongoing on further improving the alignment software
and strategy, in order to be well prepared once the first collisions
will be delivered by the LHC.


\end{multicols}

\begin{table}
\caption{Anticipated rates of $W^\pm\to\mu^\pm\nu$ and $Z^0\to\mu^+\mu^-$ events after HLT in 2008.}
\label{tab:zwrates}
\centering
\begin{tabular}{lccccc}
\hline
Luminosity & \multicolumn{2}{c}{$10^{32} \rm\ cm^{-2} s^{-1}$} &  \multicolumn{3}{c}{$2*10^{33} \rm\ cm^{-2} s^
{-1}$} \\
\hline
Time          &     few weeks  & 6 months  &   1 day   &  few weeks & one year \\
Int. Luminosity   &   $100 \rm\ pb^{-1}$ & $1 \rm\ fb^{-1}$ &   &  $1 \rm\ fb^{-1}$ &   $10 \rm\ fb^{-1}$ \\
\hline
\hline
$W^\pm\to\mu^\pm\nu$ & 700K & 7M & 100K & 7M & 70M \\
\hline
$Z^0\to\mu^+\mu^-$   & 100K & 1M &  20K & 1M & 10M \\
\hline
\end{tabular}
\end{table}

\begin{figure}
\centering
\includegraphics[width=0.4\linewidth]{Figure_003-012-a.eps}
\includegraphics[width=0.4\linewidth]{Figure_003-012-b.eps} 

\caption{Muon Alignment: $r\phi$ estimator for muon chambers in different wheels for 
aligned (solid line) and $\pm1\rm\ mm$ displaced samples of $W^\pm$.
}
\label{fig:muonali}
\end{figure}

\end{document}